\documentstyle[12pt,aasms4,psfig]{article}

\def\simlt{\lower.5ex\hbox{$\; \buildrel < \over \sim \;$}}
\def\simgt{\lower.5ex\hbox{$\; \buildrel > \over \sim \;$}}

\def\gcm3{{\rm\,g\,cm^{-3}}}
\def\ncm3{{\rm\,cm^{-3}}}

\def\>{$>$}
\def\<{$<$}

\lefthead{Crocker, Melia \& Volkas}
\righthead{Sgr A East Neutrinos}

\begin{document}
\title{ 
%\rightline{November 1999}
%\rightline{{\tt UM-P-99/40}}
%\rightline{{\tt RCHEP-99/08}}
%\ \\
Oscillating Neutrinos from the Galactic Center}

%\title{The Role of Magnetic Field Dissipation in the\\
%       Black Hole Candidate Sgr A*}

%\author{Robert F. Coker\altaffilmark{1}$^*$ 
%and Fulvio Melia\altaffilmark{2}$^{*\dag}$}
%\affil{$^*$Physics Department, The University of Arizona, Tucson, AZ 85721}
%\affil{$^{\dag}$Steward Observatory, 
%The University of Arizona, Tucson, AZ 85721}

\author{Roland M. Crocker$^*$, 
Fulvio Melia\altaffilmark{1}$^{\dag}$ 
and Raymond R. Volkas$^*$}
\affil{$^*$School of Physics,
Research Centre for High Energy Physics,
\\The University of Melbourne,
3010 Australia\\
r.crocker, r.volkas@physics.unimelb.edu.au}
\affil{$^{\dag}$Physics Department and Steward Observatory, 
\\The University of Arizona, Tucson, AZ 85721\\
melia@physics.arizona.edu}

\altaffiltext{1}{Presidential Young Investigator}

%------------------------------------------------------------------------------
%\documentstyle[epsfig,12pt,preprint,tighten,aps]{revtex}
%\begin{document}

%\title{ \rightline{November 1999}
%\rightline{{\tt UM-P-99/40}}
%\rightline{{\tt RCHEP-99/08}}
%\ \\
%Oscillating Neutrinos from the Galactic Center}

%\author{R. M. Crocker$^{(a)}$, 
%F. Melia$^{(b)}$\footnote{Presidential Young Investigator} 
%and R. R. Volkas$^{(a)}$}
%\vspace{0.6cm}
%\address{$^{(a)}$School of Physics\\
%Research Centre for High Energy Physics\\
%The University of Melbourne\\
%3010 Australia\\
%r.crocker, r.volkas@physics.unimelb.edu.au}

%\vspace{0.6cm}
%\address{$^{(b)}$
%Physics Department and Steward Observatory\\
%The University of Arizona\\
%Tucson AZ 85721	
%United States\\
%melia@physics.arizona.edu}

%\maketitle

%-----------------------------------------------------------------------------

\begin{abstract}

It has recently been demonstrated that the $\gamma$-ray emission 
spectrum of the EGRET-identified,
central Galactic source 2EG J1746-2852 can be well 
fitted by positing that these photons are generated by the decay of 
$\pi^0$'s produced in p-p scattering at or near an energizing shock. 
Such scattering also produces charged pions which decay leptonically.
The ratio of $\gamma$-rays to neutrinos generated by the central Galactic
source may be accurately determined 
and a well-defined and potentially-measurable 
high energy neutrino flux at Earth is unavoidable.
%This flux is above the atmospheric neutrino background for $E_{\nu}> 
%\text{few} 
%\times TeV$ (assuming a neutrino telescope angular resolution of $\sim 
%2^{\circ}$). 
An opportunity, therefore, to detect neutrino oscillations over an
unprecedented scale is offered by this source. In this paper we assess the
prospects for such an observation with the generation of neutrino \v{C}erenkov 
telescopes now in the planning stage. We determine that the next generation of 
detectors may well find an oscillation signature in the Galactic Center (GC)
signal.
%, but that such an observation will 
%probably not further constrain the oscillation
%parameter space mapped out by current atmospheric, solar, reactor and
%accelerator neutrino oscillation experiments. 
%Strong confirmation of the
%evidence will require a further generation of detectors and/or techniques
%that can determine the presence of $\nu_e$ flux at lower energies than
%it is currently predicted will be available for interrogation
%by the next generation of
%detectors.
\end{abstract}


\section{Introduction}

\subsection{The Neutrino Source}

The dominant radio emitting structures at the Galactic Center (GC) are   
the supernova remnant (SNR)-like shell Sagittarius (Sgr) A East, 
a three-armed spiral of ionized gas dubbed Sgr A West, and, 
embedded at the center of Sgr A West, the Galactic dynamical nucleus, Sgr A*, 
thought to be a massive ($M \simeq 2.6 \times 10^6 M_{\odot}$) 
black hole (\cite{haller}; \cite{genzel}; \cite{ghez}).
Sgr A East has a major axis length of about 10.5 pc and its 
center is located 2.5 pc from Sgr A* in projection,
and probably behind the latter (\cite{goss}). 
Lo et al. (1998) have recently determined 
the intrinsic size of Sgr A* to be less 
than $5.4 \times 10^{11}$ m at $\lambda 7$mm. 

Also located at the GC is the EGRET-identified 
$\gamma$-ray source 2EG J1746-2852 (\cite{mayer}).
It has been shown that the high energy ($0.1 - 10 GeV$) 
$\gamma$-ray emission spectrum 
of this source is very likely due to the decay of $\pi^0$'s
 (\cite{melia}; \cite{markoff}). 
These pions are produced by p-p collisions which might 
plausibly take place at 
either of two shock regions: 1) the shock at Sgr A* due to gas 
accretion from ambient winds, or 2) the shock produced by the expansion of 
the SNR-like nonthermal shell 
of Sgr A East into the ambient gas of the interstellar
medium. Thus, {\it a priori}, either or both Sgr A* and Sgr A East 
might be the source of the 
$\gamma$-rays which constitute 2EG J1746-285. It has recently been 
shown, however, that the identification of
Sgr A* with 2EG J1746-28 is disfavored because charged leptons
produced in $\pi^\pm$ decays would emit too much synchrotron flux 
in Sgr A*'s intense magnetic field 
at GHz frequencies to be consistent with the well-studied radio spectrum
of this object (\cite{melia}; \cite{blasi}).   

On the other hand, given the physical conditions in 
Sgr A East, the putative charged leptons generated there
have a distribution that mimics a power-law with index $\sim 3$.
The synchrotron flux radiated by these charges is
consistent with the radio spectrum of Sgr A East observed with the VLA.
In fact, such relativistic electrons and positrons would
also radiate by bremsstrahlung and undergo 
inverse Compton scattering in such a way
as to self-consistently explain the
entire broadband emission spectrum of Sgr A East, ranging 
from GHz frequencies all the way up to the 
TeV energies observed by Whipple (\cite{buckley}). 
For the purposes of this paper, then, we shall take it that the 
EGRET source 2EG J1746-285 is identical with Sgr A East (\cite{melia}).
We note in passing that the 
maximum energy attained by the shocked protons
at Sgr A East, given the energy loss rate via collision in the shock,
is $\sim 5 \times 10^{15} \ eV \ = \ 5000 \ TeV$ (\cite{melia}).

Regardless of the ultimate identity of the EGRET source 2EG J1746-285, 
given that the process producing the high energy emission is pionic,
there should be an associated neutrino flux from the
GC (\cite{blasi}). 
These neutrinos are due both to direct pion decay ($\pi^\pm \rightarrow 
\mu \nu_{\mu}$) and to the decay of 
 muons to electrons and positrons ($\mu^\pm \rightarrow e 
\nu_e \nu_\mu$), where we take $\nu$ to mean $\nu$ and $\overline{\nu}$
here (as we shall often do in the remainder of this paper). 
{\it Prima facie}, then, we expect the flavor 
composition of the neutrino `beam' 
 generated at the GC to be essentially
 $67\%$ $\mu$-like
and $33\%$ $e$-like by na\"{\i}ve 
channel counting (c.f. atmospheric neutrinos in the GeV energy range). 
Note that there is a $\nu_{\tau}$ background produced at the 
source due to non-pionic
processes like charmed hadron decay. This background is, however, small;  see
later. Of course, in the absence of neutrino flavor
oscillations, one would expect to observe G.C. neutrinos at Earth with
the same flavor composition as that generated at the source. 

We do not distinguish between $\nu$
and $\overline \nu$, because present and planned terrestrial
detectors do/will not distinguish between the two. There is one
interesting proviso to this statement, however: a
$\overline \nu_e$ flux at $E_{\overline\nu_e} 
\simeq 6.4 \times
10^3 \ TeV \ = \ 6.4 \times 10^{15} \ eV$ 
can be detected by resonant $W^-$ boson
production via $\overline \nu_e e^{-} \to W^{-}$ with
the electrons in the detector medium. The resonance energy, however, is just
above that attained by neutrinos generated in the processes described
above at the GC (\cite{glashow}; \cite{berezinsky}; \cite{gandhi1}).

Given our detailed knowledge of the basic physical processes
producing the GC $\gamma$-rays, we are able to 
determine an expression for the total neutrino 
emission at the source, $Q_{\nu}(E_{\nu})$, in terms
of the $\gamma$-ray emission there, $Q_{\gamma}(E_{\gamma}^0)$,
the numerical power of the proton spectrum   
at the source, $\alpha$
(such as would result from shock acceleration at either Sgr A East 
or Sgr A*), and $r \equiv ({m_\mu}/{m_\pi})^2$ (\cite{blasi}). The quantity
$\alpha$ has been empirically-determined to lie between 2.1 and 2.4
 (\cite{markoff}; \cite{blasi}), using a procedure to fit the EGRET spectrum
of 2EG J1746-2852 with a detailed calculation of the particle
cascade using an extensive compilation of pion-multiplicity cross-sections. 
In the energy range between the $\Delta$-resonance ($\sqrt{s}\sim 1$ GeV) and
the ISR (Intersecting Storage Rings) range ($\sim 23-63$ GeV), simple 
scaling (\cite{feynman}) does not adequately take into account the 
strong dependence of the cross section on the rapidity at lower energy,
and the pion distribution is not adequately described by a
power-law mimicking the injected relativistic proton distribution
between $\sim 1$ and $\sim 100$ GeV. Instead, the distribution steepens
in this region and is curved, which is consistent with the suggested
spectral shape measured by EGRET. Above about $10$ GeV, however, the
pion distribution settles into the `asymptotic' form suggested by
scaling, where the power-law index is a direct reflection of the
underlying relativistic protons.  Thus, an EGRET spectrum with an
effective spectral index of $\sim -3$ below $10$ GeV is produced by
a pion distribution whose power-law index lies in the range $2.1-2.4$
above this energy.  In other words, a relatively steep and curved $\gamma$-ray 
spectrum below $10$ GeV is consistent with a flatter neutrino
spectrum at TeV-energies.  The relative normalization between the
$\gamma$-ray and neutrino distributions is effected at $10$ GeV where
the pions take on a power-law form. 

We take the neutrino spectrum at Earth to be, in general, given by:
\begin{equation}
\Phi_{\nu}(E_{\nu}) = \Phi_{\nu}(10 \ GeV) \  (E_{\nu}/{10 \ GeV})^{-\alpha} 
\end{equation}
Normalizing to the observed $\gamma$-ray flux at Earth at $10$ GeV, 
one arrives at the following values for the total neutrino flux here
 (\cite{blasi}): 
\vspace{0.2cm}
\begin{equation}
\Phi_{\nu}(E_{\nu}) = 1.1 \times 10^{-9} \ (E_{\nu}/{10 \ GeV})^{-2.1} \ \ 
 cm^{-2}\, s^{-1}\, GeV^{-1}
\end{equation}
for $\alpha = 2.1$, and 

\begin{equation}
\Phi_{\nu}(E_{\nu}) = 9.6 \times 10^{-10} \  (E_{\nu}/{10 \ GeV})^{-2.4} \ \ 
cm^{-2} s^{-1} GeV^{-1}
\end{equation}
\noindent
for $\alpha = 2.4$, where we have taken the absolute upper bound to the
energy spectrum of G.C. neutrinos to be given by the highest energy 
($5 \times 10^{15} eV$) of the shocked protons. 
(Kinematical calculations show that
neutrinos created by the decay of charged pions produced in scattering of 
a `beam' proton off a stationary `target' proton can
attain energies very close to the `beam' proton.) Note that in the above we
make the very reasonable assumption 
that high energy $\gamma$'s and $\nu$'s travel
to Earth equally unimpeded by the ambient matter they encounter 
(which has a column
number density of barely $10^{23} \ cm^{-2}$). 

Two factors improve the odds for the detection of the GC neutrino flux above
the atmospheric neutrino background. These are 1) the effectively point-source
nature of the GC, and 2) a GC neutrino spectrum that is significantly
flatter than that of atmospheric neutrinos (which goes as $E_{\nu}^{-3.7}$).
If we preliminarily adopt an 
angular resolution of $\theta_{res} \sim 2^\circ$ for the
proposed large scale detectors ($1 \  km^2$ effective detector area), the
condition for the detection of the GC neutrino flux 
is $\Phi_{\nu}(E_{\nu})  / \Omega_{res} > I_{atm}(E_{\nu})$, 
where 
%$\Phi_{\nu}(E_{\nu})$ is the total neutrino flux due to the GC source,
$\Omega_{res} \approx \pi \theta_{res}^2$ is the solid angle 
corresponding to the angular resolution of the experiment 
and $I_{atm}(E_{\nu})$ is the flux of atmospheric neutrinos 
per unit solid angle. This condition is fulfilled above a few 
TeV, and the expected event rate from this preliminary analysis 
is $\sim 4 \ km^{-2} yr^{-1}$ for
$\alpha = 2.4$ to $\sim 70 \ km^{-2} yr^{-1}$ for 
$\alpha = 2.1$ (\cite{blasi}). Note
 that a fuller analysis of event rates
(presented later)
must also consider the problems posed by the atmospheric 
{\it muon} background and Earth neutrino opacity.

We see therefore that preliminary calculations reveal that there is a 
well-determined and potentially observable  neutrino flux at the Earth 
from the Galactic Center.  We now briefly list the motivations behind
this work before going on to consider whether any sort of neutrino 
oscillation signature might be detectable in the GC signal.

\subsection{Summary of Motivations}

The main motivations behind the present work are:
\begin{enumerate}
\item Sgr A East is arguably the most thoroughly understood extra-solar 
astrophysical source of very high energy neutrinos identified to date. It is 
thus of fundamental importance for the embryonic science of neutrino
astronomy.
\item It is important for general scientific reasons to explore the neutrino 
oscillation phenomenon in a wide variety of regimes. 
Because of the high energy scales and the very
long baselines involved, astrophysical sources
such as Sgr A East provide a novel regime not investigated in previous
and current solar, atmospheric, reactor 
and accelerator neutrino detection experiments. 
Previous works to have considered propagation effect 
signatures in galactic and extra-galactic high energy neutrino signals 
include 
\cite{learned,weiler,sandip,roy,wudka,enqvist,husain,saltzberg,bento,iyer,
mannheim,raffelt}. 
\item Given that solar and atmospheric neutrino observations have essentially
established the existence of neutrino oscillations, it is important 
to incorporate this propagation effect when examining possible sources for
study through neutrino astronomy. Neutrino signals from astrophysical 
sources are an important complement to electromagnetic signals from same, 
and they will serve to improve our understanding of the dynamics of important
astrophysical objects such as supernova remnants, gamma ray bursters and
active galactic nucleii.
\end{enumerate}

\section{Neutrino oscillations between Sgr A East and Earth}

\subsection{Distance Considerations}

For purposes of calculational expediency we take the neutrino 
source Sgr A East to have a linear dimension of
$10\ pc \simeq 3 \times 10^{17}\ m$.
%The alternative source, Sgr A* we shall take to have dimension
%$5  \times 10^{11}\ m$. 
This distance is relevant because we need to know how the neutrino
oscillation lengths compare with the size of the emitting object to
determine whether the neutrino source is flavor coherent.
If the former are small compared to the latter, then,
because neutrinos are emitted from all points within the source,
the oscillations will be averaged out. Alternatively, if the
latter are large compared to the former, then no averaging due
to the finite size of the source will be needed 
and the source is essentially
flavor coherent for neutrinos of a given energy.
Note that two
types of averaging generally need to be done: over distance, and
over energy. Thus far we have only considered distance 
averaging due to the finite size of the $\nu$ source. One also has
to take into account distance (and energy) averaging due
to the detector. For Sgr A East the source distance 
scales involved are at least
six orders of magnitude larger than those for the detector (1 A.U. $\simeq
1.5 \times 10^{11}$ m). Detector-based
distance averaging, then, will not impact on calculations concerning 
Sgr A East. 
%For Sgr A*, however, the source size is comparable to 1 A.U. and 
%distance averaging at both detector and source becomes important. 
We do not address the issue of energy averaging due to
the finite energy resolution of the detector in great detail in this paper.

The distance between source and detector is about
\begin{equation}
8\ kpc \simeq 2.5 \times 10^{20}\ m.
\end{equation}

%According to Blasi and Melia, the expected $\nu$ flux from
%this source is above the atmospheric $\nu$ background for
%energies
%\begin{equation}
%E > 3\ TeV.
%\end{equation}
%This defines the ``interesting'' energy range. It is certainly
%the correct energy range for proposed neutrino telescopes.

\subsection{Introduction to Neutrino Oscillations}

We consider only 2-flavor oscillation modes $\nu_\alpha 
\leftrightarrow \nu_\beta$ for simplicity and definiteness.
Suppose a beam of flavor $\alpha$ is produced at $x = 0$.
Then at a point $x$ distant from the source the oscillation
probability is
\begin{equation}
P(\alpha \to \beta) = \sin^2 2\theta \sin^2 \left(
\pi \frac{x}{L} \right),
\end{equation}
whereas the ``survival probability'' is obviously
\begin{equation}
P(\alpha \to \alpha) = 1 - P(\alpha \to \beta).
\end{equation}

The parameter $\theta$
is the `mixing angle' which determines
the amplitude of the oscillations. 
The value $\theta = {\pi}/{4}$, which leads to the largest possible
amplitude, is termed `maximal mixing'.
The parameter $L$
is the `oscillation length' and is given by
\begin{equation}
L = \frac{4 \pi E}{\Delta m^2}
\end{equation}
in natural units $\hbar = c = 1$. Note
that the oscillation length increases linearly
with energy. This is important because
the high energy scale under consideration $(E > TeV)$
stretches the oscillation length.
The parameter
$\Delta m^2 \equiv |m_1^2 - m_2^2|$ is the
squared-mass difference between the two mass 
eigenstate neutrinos.

Totally averaged oscillations see the second $\sin^2$
factor in Equation (5) set equal to ${1}/{2}$, leading to
\begin{equation}
\langle P(\alpha \to \beta) \rangle = \frac{1}{2}
\sin^2 2\theta.
\end{equation}
This, to reiterate, can be due to either distance or
energy spread or both.

Given the poor statistics of the proposed
neutrino telescopes, only modes with large mixing angles,
$\theta$, can be probed (unless the MSW phenomenon takes
place -- see later). The atmospheric neutrino
anomaly (for $\nu_\mu$'s) 
seen by SuperKamiokande and other experiments
clearly indicates large angle vacuum oscillations,
however (\cite{fukuda1}; \cite{apollonio}).
 Further, the solar neutrino anomaly 
(for $\nu_e$'s) can
be solved by large angle oscillations (or by small
angle oscillations through the MSW effect) (\cite{smy}).
 In summary, then,
the atmospheric anomaly {\it definitely} requires a large
mixing angle solution, while the solar problem {\it can}
be solved by large angle oscillations.

We now briefly review the various possible solutions to
the atmospheric and solar neutrino problems, and then
apply the various scenarios to the GC neutrino
flux.

\subsection{Atmospheric Neutrinos}

SuperKamiokande detects a $50\%$ deficit of $\mu$-like atmospheric neutrinos
coming up through the Earth (\cite{fukuda1}).
 They see no deficit of either
upward- or downward-going 
$e$-like neutrinos. The lower energy downward-going $\mu$-like
events are deficient, whereas their high-energy counterparts are not.
These data can be explained by close-to-maximal $\nu_{\mu} \to \nu_x$
oscillations with $x \neq e$ and $x = \tau$ or $x = s$ (sterile).
These two alternatives both require parameters in the range: 

\begin{equation}
\nu_\mu \to \nu_x \quad {\rm with}\ \Delta m^2_{\mu x} =
10^{-3} \to 10^{-2}\ eV^2 \quad {\rm and}\quad 
\sin^2 2\theta_{\mu x} = 1.  \label{mutau}
\end{equation}

%\begin{eqnarray}
%& \nu_\mu \to \nu_x \quad {\rm with}\ \Delta m^2_{\mux} =
%10^{-3} \to 10^{-2}\ eV^2 \quad {\rm and}\quad 
%\sin^2 2\theta_{\mux} = 1 & \label{mutau}\\
%& \nu_\mu \to \nu_s \quad {\rm with}\ \Delta m^2_{\mu s} =
%10^{-3} \to 10^{-2}\ eV^2 \quad {\rm and}\quad 
%\sin^2 2\theta_{\mu s} = 1 &.\label{mus}
%\end{eqnarray}
\noindent
(To be strict, the $\Delta m^2$ ranges are a little
different for the two possibilities because
of the `matter effect' in the Earth, but this will
be irrelevant for us  (\cite{foot}; \cite{atmos}).) 
SuperKamiokande currently favors oscillations to
$\nu_\tau$ over oscillations to a sterile neutrino at the 2$\sigma$ level
(though this is a very preliminary result) (\cite{takita}).

\subsection{Solar Neutrinos}

The solar neutrino problem can be solved by $\nu_e \to \nu_y$
oscillations, where $y = \mu, \tau, s$ are all allowed, with
one important proviso: if the Los Alamos LSND experiment is
correct, then $\nu_e \to \nu_\mu$
oscillations, with parameters that cannot solve the solar
neutrino problem, have already been detected (\cite{LSND}).
So, if the still-controversial LSND result is correct, 
then $y = \mu$
is ruled out. The MiniBOONE and BOONE
experiments at Fermilab should eventually settle 
this issue (\cite{bazarko}).

The precise oscillation parameter space required to account for
the solar data depends on which of the solar neutrino
experiments are held to be correct. The two parameter ranges defined
 below, however, are broadly consistent with all solar data; 

%The experiments are as
%follows:
%\begin{enumerate}
%\item GALLEX, low-E threshold, rate only, 
%radiochemical, calibrated
%by external $\nu_e$ source. They see roughly a $50\%$
%deficit. They have no directional information.
%\item SAGE, same as for GALLEX.
%\item Homestake, mid-E threshold, rate only, radiochemical, 
%never been calibrated by external $\nu_e$ source,
%oldest experiment, no directional information.
%They see roughly a $65\%$ deficit.
%\item Kamiokande, high-E threshold, rate only,
%real-time, have directional information. They
%see roughly a $50\%$ deficit.
%\item SuperKamiokande, high-E threshold, rate + 
%day-night asymmetry + energy spectrum + seasonal
%variation, real-time, have directional information.
%See roughly a $50\%$ deficit.
%\end{enumerate}

%The interpretation of these results is not as
%straightforward as for atmospheric neutrinos. Not
%all of the claimed results need be correct. It's
%a long story to go through all of the possibilities.
%In a nutshell, I would say the following:
\begin{enumerate}
\item $\nu_e \to \nu_y$ with a {\it small} mixing
angle (SMA) $\theta_{ey}$ is possible through the MSW
effect. If this pertains, then the oscillation
amplitude will be far too small to affect Sgr A East
neutrinos.
\item $\nu_e \to \nu_y$ with a very large mixing angle
(LMA) $\sin^2 2\theta_{ey} \simeq 1$ is an interesting
possibility for the range
\begin{equation}
10^{-3} \ga \Delta m^2_{ey}/eV^2 \ga 10^{-10}.
\end{equation}
\end{enumerate}
The immediate vicinity of $\Delta m^2_{ey} \sim 10^{-10}\ eV^2$ 
%(down to $\sim 4 \times 10^{-10}$) 
defines
`just-so' oscillations where the oscillation
length for solar neutrinos 
is of order 1 A.U. $ \ $For larger $\Delta m^2_{ey}$
values completely averaged oscillations,
with a flux suppression factor of $0.5\sin^2 2\theta_{ey}$, result.
Maximal mixing explains almost all
of the data with averaged oscillations (excepting
 the Homestake result (\cite{cleveland}), 
and the controversial SuperK spectral
anomaly). Values of $\Delta m^2_{ey} > 10^{-3}\ eV^2$ are ruled out by
the non-observation of $\overline \nu_e$ disappearance
from reactors (CHOOZ, Palo Verde experiments (\cite{chooz}; \cite{boehm})).

\subsection{Atmospheric and Solar Neutrino Data Combined}

In summary, for GC neutrinos the following are
well motivated scenarios that are composed of
2-flavor subsystems:
\begin{enumerate}
\item Large angle $\nu_e \to \nu_s$ + large angle
$\nu_\mu \to \nu'_s$ (scenario 1).\footnote{This is the situation 
predicted by the Mirror Matter or 
Exact Parity Model. See  (\cite{mirror1}; \cite{mirror3}; \cite{mirror4}).} 
\item Large angle $\nu_e \to \nu_s$ + large angle 
$\nu_\mu \to \nu_\tau$ (scenario 2).
\item Large angle $\nu_e \to \nu_\tau$ + large angle $\nu_\mu 
\to \nu_s$ (scenario 3).
\item Small angle $\nu_e \to \nu_y$ + large angle $\nu_\mu \to \nu_s$
(scenario 4).
\item Small  angle $\nu_e \to \nu_y$ + large angle $\nu_\mu 
\to \nu_\tau$ (scenario 5).
\end{enumerate}

We are now in a position to perform a number of 
simple calculations for neutrino oscillations
between the GC and the Earth motivated by the above list of
2-flavor
possibilities. Note here that bimaximal
 (\cite{bimax1}; \cite{bimax2};
\cite{bimax3}; \cite{bimax4}; \cite{bimax5}; \cite{bimax6})
and trimaximal  (\cite{trimax1}; 
\cite{trimax2}; \cite{trimax3})
mixing scenarios, which are intrinsically 3-flavor, 
will not be considered in this paper.

Using the atmospheric problem parameters, we see that the 
$\nu_\mu \to \nu_x$ oscillation length is given by:
\begin{equation}
L_{\mu x} \simeq 2.5 \times 10^{8}
\frac{E/(1\ TeV)}{\Delta m^2_{\mu x}/(10^{-2}\ eV^2)}\ m.
\end{equation}
Therefore, the oscillation length is orders of magnitude less
than the size of Sgr A East for the entire neutrino spectrum (which only 
reaches up to 
$5 \times 10^{15} \ eV = 5 \times 10^{3} \ TeV$).
% less  except for very high energies
%\begin{equation}
%E > 10^{9}\ TeV.
%\end{equation}
This means that the oscillations will be distance averaged,
and hence at Earth we expect
a $50/50$ mixture of $\nu_\mu$
and $\nu_x$, where $x = \tau$ or $x = s$ depending on which
solution to the atmospheric problem turns out to be the correct
one. 
%Alternatively, for Sgr A*, the source size is smaller than the
%oscillation length for 
%\begin{equation}
%E > 2 \times 10^{3}\ TeV.
%\end{equation}
%Note, further, that energy averaging due to the finite
%energy resolution of the terrestrial detectors must also be considered.

Using the solar problem parameters one determines 
the $\nu_e \to \nu_y$ oscillation length to be
\begin{equation}
L_{e y} \simeq 2.5 \times 10^{15}
\frac{E/(1\ TeV)}{\Delta m^2_{ey}/(10^{-9}\ eV^2)}\ m.
\end{equation}
The reference $\Delta m^2_{ey}$ is in the `just-so' range.
The oscillation length of $\nu_e \to \nu_y$ oscillations in this range,
therefore, becomes larger than Sgr A East
for $E > 10 - 100\ TeV$ or so.
This means that the more energetic component of the
$\nu_e$ beam from the source is flavor-coherent. 

In principle, such coherence
would evidence itself by an energy dependent spectral distortion; 
the $\nu_e$ flux 
at a particular energy ($E \to E+\Delta E$) 
would depend on the part of the neutrino
oscillation wave (for that particular energy) encountered by the 
Earth at its distance from Sgr A East, 
i.e. the neutrino flux at a particular energy might be anything
from maximally suppressed to unsuppressed depending exactly
on $\Delta m^2_{ey}$ and the source-observation point distance.
Certainly, ranging over the expected energy spectrum
(and therefore ranging over $L_{ey}$), we
should see the flux vary (over and above the variation given by
the spectral shape) between  maximally suppressed and unsuppressed.
Imagining, then,
that we had both a neutrino detector able to determine the 
energy of an incoming neutrino to arbitrary accuracy,
and that we had a very long time to accumulate statistics,
we should be able to find an experimental signature of the 
flavor-coherence in the form of this spectral distortion
(and thus determine whether $\Delta m^2_{ey}$
were in the `just-so' energy range 
able to lead to such coherence, and, if it were, 
exactly what value it takes). 
%. flux in a small
%energy bin might be anything from maximally suppressed to unsuppressed.
%would vary from being maximally suppressed to unsuppressed
%between one detector energy `bin' and another, given a sufficiently small
%bin size. 
Pragmatically, given the small statistics 
that will accrue from the GC source and the limited energy 
resolution expected to be achieved by any of 
the proposed neutrino telescopes, 
one expects no observational consequence of the flavor coherence. 
This is because the energy dependence of the flux suppression
washes out with the inevitably large size of the energy bins
particular neutrino events are accumulated into.
The beam, therefore, 
is indistinguishable from one in the distance-averaged 
oscillation regime.

Note also that 
the $\nu_e \to \nu_y$ oscillation length
would become of the order of the GC-Earth distance 
for $E \sim 10^{16}\ eV$
for $\Delta m^2_{ey} = 10^{-9}\ eV^2$. The $\nu_e$ flux would, then,
rise from being suppressed below $10^{17}\ eV$ to unsuppressed
above $10^{17}\ eV$ if the $\nu_e$ attained this energy. 
Of course given that
the maximum energy of the shocked protons does not surpass  
$\sim 5 \times 10^{15}\ eV$ 
this phenomenon does not occur for the GC source.

At the opposite extreme of the acceptable parameter space, i.e., 
$\Delta m^2_{ey} \simeq 10^{-3}\ eV^2$, the oscillation length is
\begin{equation}
L_{e y} \simeq 2.5 \times 10^{9}
\frac{E/(1\ TeV)}{\Delta m^2_{ey}/(10^{-3}\ eV^2)}\ m.
\end{equation}
This is back in the totally distance-averaged oscillation regime. 
In conclusion,
for the entire allowable $\Delta m^2_{ey}$ regime we pragmatically expect 
a situation similar to the muon-type neutrino case: 
totally averaged oscillations, 
i.e., a $50/50$ mixture of $\nu_e$ and $\nu_y$ for maximal mixing.

%I've only considered totally averaged oscillations
%in this list. Given the poor statistics of
%the proposed experiments, presumably the $50\%$
%reduction cannot be established. Therefore, the
%``minimal science'' will be:
%\begin{enumerate}
%\item Positive detection of {\it some} neutrinos
%from the GC.
%\item Existence or not of a $\nu_\tau$ component
%from the GC, presuming that the double-bang
%signature of Learned-Pakvasa can be utilized.
%\end{enumerate}

\subsection{Matter Effects?}

A brief calculation is sufficient to show that for the GC, matter effects
(refractive indices for neutrinos) do not impinge significantly on the
oscillation probabilities.  The quantities that have to be compared are
${\Delta m^2}/{2E}$ and  $\sim G_F n$,
where $G_F$ is the Fermi constant, and $n$ is the electron minus
positron number density for the medium. 
Right at the source we expect $n \sim 0$ because of the equal production
of electrons and positrons there. 
Concerning propagation of the neutrinos from source to detector, we have that 
the interstellar medium consists of approximately 1 H atom per cm$^3$ so that
$G_F n \simeq (10^{-5}\ GeV^{-2})(2 \times 10^{-14}\ GeV)^3$,
converting the number density to natural units.
This number works out to be about $10^{-46}\ GeV$.
The smallest $\Delta m^2$ we consider is $10^{-10}\ eV^2 =
10^{-28}\ GeV^2$ and for the highest attainable neutrino
 energy of $5 \times 10^{15} \ eV = 
5 \times 10^{6}\ GeV$, we get $\Delta m^2/E \sim 2 \times 10^{-35}\ GeV$,
so we are 11 orders of magnitude away from having
important matter effects due to the interstellar medium.
We do not consider matter effects due to dense intervening
objects between the GC and Earth, since their covering fraction
for Sgr A East is trivially negligible.

\subsection{Observational Consequences -- in Theory}
We consider now the observational consequences of scenarios 1 to 5 
listed above in terms of the neutrino flux at Earth (we remind the reader
that all $\nu_\mu$ mixing scenarios are LMA). 

\begin{enumerate}
\item Scenario 1 (LMA $\nu_e \to \nu_s$ and $\nu_\mu \to \nu'_s$): 
$50\%$ reduction of both $\nu_e$ and 
$\nu_\mu$ flux, and no $\nu_\tau$ appearance above 
background.

\item Scenario 2 (LMA $\nu_e \to \nu_s$ and $\nu_\mu \to \nu_\tau$): 
$50\%$ reduction of $\nu_e$ flux,
and equal $\nu_\mu$ and $\nu_\tau$ fluxes.

\item Scenario 3 (LMA $\nu_e \to \nu_\tau$ and $\nu_\mu 
\to \nu_s$): Equal $\nu_e$ and $\nu_\tau$
fluxes, and $50\%$ reduction of $\nu_\mu$ flux.

\item Scenario 4 (SMA $\nu_e \to \nu_y$ and $\nu_\mu \to \nu_s$): 
Unreduced $\nu_e$ flux, $50\%$
reduced $\nu_\mu$ flux, no $\nu_\tau$ appearance above background.

\item Scenario 5 (SMA $\nu_e \to \nu_y$ and $\nu_\mu 
\to \nu_\tau$): Unreduced $\nu_e$ flux, and
equal $\nu_\mu$ and $\nu_\tau$ fluxes.
\end{enumerate}

The scenarios above imply the following ratios (and ratios of ratios) of
neutrino flavor fluxes:

\vspace{0.5cm}
\begin{tabular}{ccccccc} \hline
Ratio & No Oscillations & Scenario 1 & 
Scenario 2  & Scenario 3 & Scenario 4  & Scenario 5 \\ 
\hline
\vspace{0.1cm}
$({\Phi_{\nu_e}^{obs}}/{\Phi_{\nu_e}^{theor}})$ & 1 & $\frac{1}{2}$ & 
$\frac{1}{2}$ & $\frac{1}{2}$ & 1 & 1 \\
\vspace{0.1cm}
$({\Phi_{\nu_{\mu}}^{obs}}/{\Phi_{\nu_{\mu}}^{theor}})$ & 1 &
$\frac{1}{2}$ & 
$\frac{1}{2}$ & $\frac{1}{2}$ & $\frac{1}{2}$ & $\frac{1}{2}$ \\ 
\vspace{0.1cm}
%$\frac{\Phi_{\nu_{\tau}}^{obs}}{\Phi_{\nu_{\tau}}^{theor}}$ & 1 & $\gg 1$ & 
%$\gg 1$ & 1 & $\gg 1$  \\
%\vspace{0.1cm}
$({\Phi_{\nu_e}^{obs}}/{\Phi_{\nu_\mu}^{obs}})$ & $\frac{1}{2}$
& $\frac{1}{2}$& $\frac{1}{2}$& $\frac{1}{2}$&1&1\\
\vspace{0.1cm}
$({\Phi_{\nu_\tau}^{obs}}/{\Phi_{\nu_\mu}^{obs}})$ & $\ll 1$
& $\ll 1$& 1 & $\frac{1}{2}$& $\ll 1$ &1\\
\vspace{0.1cm}
$\frac{{(\Phi_{\nu_e}}/{\Phi_{\nu_{\mu}})^{obs}}}
{({\Phi_{\nu_e}}/{\Phi_{\nu_{\mu}}})^{theor}}$ 
& 1 & 1 & 1 & 1 & 2 & 2\\ 
\vspace{0.1cm}
$\frac{{(\Phi_{\nu_{\tau}}}/{\Phi_{\nu_{\mu}})^{obs}}}
{({\Phi_{\nu_{\tau}}}/{\Phi_{\nu_{\mu}}})^{theor}}$ 
&1 & 2 & $\gg 1$ & $\gg 1$ & 2 & $\gg 1$ \\
\vspace{0.1cm}
$\frac{{(\Phi_{\nu_e}}/{\Phi_{\nu_{\tau}})^{obs}}}
{({\Phi_{\nu_e}}/{\Phi_{\nu_{\tau}}})^{theor}}$ 
& 1 & $\frac{1}{2}$  & $\ll 1$ & $\ll 1$ & 1 & 
$\ll 1$\\ \hline
\end{tabular}

\vspace{0.5cm}

%\begin{eqnarray}
%& \frac{\Phi_{\nu_e}^{obs}}{\Phi_{\nu_e}^{theor}}\\
%& \frac{\Phi_{\nu_{\mu}}^{obs}}{\Phi_{\nu_{\mu}}^{theor}}\\
%& \frac{\Phi_{\nu_{\tau}}^{obs}}{\Phi_{\nu_{\tau}}^{theor}}\\
%& \frac{{(\Phi_{\nu_e}}/{\Phi_{\nu_{\mu}})^{obs}}}
%{({\Phi_{\nu_e}}/{\Phi_{\nu_{\mu}}})^{theor}}\\
%& \frac{{(\Phi_{\nu_{\tau}}}/{\Phi_{\nu_{\mu}})^{obs}}}
%{({\Phi_{\nu_{\tau}}}/{\Phi_{\nu_{\mu}}})^{theor}},\\
%\end{eqnarray}

\noindent 
The superscript `$obs$' denotes the flux ratios 
observed by a neutrino telescope,  while
`$theor$' denotes the ratio expected from the 
no-oscillation theoretical calculation. 
Deviation away from the value predicted for the
no oscillation case in any of the ratios defined above, 
beyond experimental uncertainty, would constitute a {\it prima facie} case for
whatever neutrino oscillation scenario most closely predicts the experimental
fluxes. Deviation in the third last ratio would constitute the strongest
evidence for oscillation because errors due to uncertainties 
in the determination of the total theoretical neutrino flux
tend to cancel in taking the ratio of the theoretical $\nu_e$ and $\nu_\mu$
flavor ratios given that $\nu_e$'s and $\nu_\mu$'s are produced by the same 
mechanism at the source. 

On the other hand,
there is considerable uncertainty concerning the $\nu_\tau$ background (see 
later) so that estimates of $\Phi_{\nu_{\tau}}^{theor}$ 
%and hence the 
%third ratio above, 
%$\frac{\Phi_{\nu_{\tau}}^{obs}}{\Phi_{\nu_{\tau}}^{theor}}$, 
may not be
particularly meaningful. For this reason, we do not list 
$({\Phi_{\nu_{\tau}}^{obs}}/{\Phi_{\nu_{\tau}}^{theor}})$.
As displayed above, though, in the absence of oscillations 
we still expect the $\nu_{\tau}$ flux to be considerably
smaller than the other flavor fluxes in the absence of oscillations
to this flavor type.
%In fact, it may be more useful to consider 
%$\frac{\Phi_{\nu_{\tau}}^{obs}}{\Phi_{\nu_{\mu}}^{theor}}$, which in the
%absence of oscillations we expect to be $\ll  1$. 
Further, deviation from 1 in the
first two ratios defined could only provide strong evidence of oscillations
if the uncertainties in the power of the neutrino spectrum, $\alpha$, and
$\Phi_{\nu}(10 \ GeV)$ were both significantly 
reduced by future $\gamma$-ray observations
using instruments with better energy resolution and coverage.  
The GLAST mission
may be the first to provide the necessary improvements 
over the next few years (\cite{glast}).

Note also that, unfortunately, none of the five scenarios 
considered here realistically 
exhibits the energy-dependent flux suppression (within an appropriate energy
range) that would be the most telling signature of neutrino oscillations.   
Further, even assuming that we possess a detector with near
perfect neutrino identification capability, so that we
 can determine the ratios defined above and hence 
distinguish between the five broad scenarios,
we still
 cannot further pin down  $\Delta m^2_{\mu x}$ or $\Delta m^2_{ey}$ than
has already been achieved with the terrestrial solar, atmospheric, 
reactor and accelerator
neutrino experiments.\footnote{As noted previously, however, 
with perfect energy resolution the potential flavor coherence of
 Sgr A East over at least some of the $\Delta {m_{ey}}^2$ parameter
space {\it would} have an experimental signature.} The  allowable
 mixing angle
parameter space might only be constrained in the sense that the above ratios
distinguish between a large and a small $\theta_{ey}$.

In the next section we examine the prospects for 
determining the neutrino flux 
of each flavor at Earth.

\section{Detection of Oscillations}

\subsection{The Detectors}

In this work we consider only the \v{C}erenkov neutrino telescopes 
now in planning and construction stages as observation platforms. Other
proposed astronomical neutrino detection methods 
tend to require
neutrino energies in excess of that possessed by 
Sgr A East 
neutrinos (see appendix C of  (\cite{rachen}) for 
a brief review).\footnote{The two most 
interesting alternative 
neutrino detection
techniques are the use of air shower arrays and 
radio detection of neutrino interactions in ice. Air shower arrays,
which probably offer the best hope for $\nu_e$ detection and identification,
are limited to energies in excess of $\sim 10^{17} \ eV$ by the 
atmospheric background (\cite{capelle}).
Radio detection of neutrinos will probably require
energies at or in excess of the GC neutrino energy upper limit 
(i.e., $5 \times 10^{15} \ eV$) because of signal-to-noise problems
 (\cite{gaisser}; \cite{alvarez1}; \cite{alvarez2}).}

The \v{C}erenkov detectors are planned to operate through the 
instrumentation of very large volumes ($\sim 1km^3$ is thought to be
optimal for astronomical neutrino detection  (\cite{halzen})) 
of some transparent medium (in practice water or ice) with photomultiplier 
tubes (PMTs). These tubes will detect the \v{C}erenkov light generated by 
superluminal charged leptons traversing the detector volume. The \v{C}erenkov
light is generated at a characteristic
angle (for the medium) away from the direction of travel 
of the charged particle. Note that
only muons and extremely energetic tauons
 have path lengths through water and ice significant on the scales
of the PMT separation of these detectors (tens of meters). Electrons
 are arrested very quickly (within a meter or so), even 
at the highest energies we are considering: $O[PeV]$.  
Lower energy tauons (produced by $\nu_\tau$
primaries with $E_{\nu_\tau} < 10^{14} \ eV)$ decay within meters. 

We remark in passing that, at 
considerably
higher energies still (i.e., $\sim 20 \ PeV$), 
the Landau-Pomeranchuk-Migdal (LPM)
effect starts to cause a measurable reduction to the pair production and
bremsstrahlung cross sections of the electron. This increases the radiation
lengths of $e^\pm$  (\cite{alvarez3}).

\v{C}erenkov neutrino telescopes of course encounter background generated by
atmospheric {\it muons} (i.e.,  muons generated directly 
by cosmic ray interactions in the 
atmosphere) as well as the atmospheric {\it neutrino} background.
In fact, this background overwhelms the genuine neutrino 
signal due to any conceivable astronomical object at sea level and 
hence neutrino telescopes must be shielded somehow. This requirement (as well
as the requirement for sufficient clarity of the medium) is what has
driven all proposed sites for working neutrino telescopes deep
(few kilometers) into the 
Antarctic icecap or underwater. 

Even at these depths,
however, the atmospheric muon background is far from negligible.
% and neutrino
%telescopes must employ some sort of triggering in order that they select only
%for genuine neutrino-generated leptons.
The simplest way to ensure exclusive selection of genuine neutrino-generated
leptons against this background 
is to have a \v{C}erenkov detector register only
 `upcoming' leptons; those that arrive at a specified angle to the vertical 
somewhat below the horizontal. It is then almost 
assured that such charged leptons have been generated by
 neutrinos which traverse some large fraction of the Earth's
diameter and then subsequently undergo charged current (CC) interactions 
with the water/ice at the detector or the rock/water/ice fairly 
close to it. 
Exclusive selection for upcoming leptons can be achieved through
a combination of geometry (simply situating all PMTs so they face 
downwards) and triggering (which can discern up-going from down-going 
signal on the basis of fairly simple 
timing considerations (\cite{AMANDA1}; \cite{AMANDA2})).
Note that the highest energy muons might traverse 
a distance of ten kilometers water equivalent and still retain
sufficient energy to produce a detectable \v{C}erenkov signal (\cite{gandhi1}).
The effective volume, therefore,
 for
$\nu_\mu$ detection is 
 substantially larger than the `instrumented' 
volume.

Of course, such a simple triggering system (and geometry)
 means that one misses out completely on the signal from
down-going neutrinos (which also generate down-going leptons
in CC interactions). This may seem like a reasonable compromise (the
effective detector area is greater from below, after all, because
of the greater amount of material below the detector than above) until one 
considers the fact that for high energy neutrinos Earth `shadowing' or 
opacity becomes a significant effect. A neutrino is shadowed when its 
interaction length becomes smaller than the distance it must travel
through the Earth to reach a detector. At energies of $\sim 10^{15} \ eV$
Earth opacity affects all neutrinos except those that reach a detector from
an almost 
horizontal direction. It would seem, therefore, that with the 
scheme described above -- reject all down-going leptons -- and the
unavoidable issue of Earth opacity, ultra high energy neutrino telescopy is
impossible, except for a tiny window on neutrinos which come from a practically
horizontal direction.

In order for ultra high energy neutrino astronomy to have a future,
detector triggering 
must be designed that does
 something smarter than simply rejecting all
down-going leptons; it must be able to select something of the genuine 
down-going signal, at least at higher energies. For the moment, 
let us assume that some more discerning
trigger can be instantiated. 
The question now is, does the GC neutrino source have a 
large enough flux at high energies to be seen against the muon background
(for reasonable values of telescope angular resolution), even
in principle? 

To make this calculation, we first adopt
 values for the high energy (vertical) 
fluxes of atmospheric muons at sea level given elsewhere 
 (\cite[fig.3]{ingelman2}) 
and then convolve these fluxes with
values for the probabilities of these muons to reach
 the particular water and ice depths of the
proposed detectors  (\cite{antonioli}) Then, to determine
event rates in a detector due to high energy atmospheric muon background,
we take these fluxes over a sensible range of telescope angular resolutions 
(from $2.0^\circ$ to
$0.3^\circ$, say). 
Subsequently, we compare these (angular resolution dependent) rates 
with the genuine, neutrino-generated event rate found by 
convolving the GC neutrino spectrum with \nopagebreak
reasonable estimates for the 
neutrino detection 
probability. 

The probability that a high
energy muon-type 
neutrino is detected in a $km^3$-scale neutrino telescope depends on
two factors, viz; approximately inversely on the interaction
length of the neutrino ($\lambda_{int}$) at 
that energy (which, in turn, depends on the charged current cross-section) 
and approximately directly on the 
radiation length of the muon ($R_\mu$) 
 produced in the interaction. 
(We assume here that the linear dimension of the detector is small
on the scale of $R_\mu$.)  
We can make a rough estimate 
of the effect of these factors by writing down a detection probability 
multiplier which goes as some power of the energy:
$$
P_{\nu \rightarrow \mu} \simeq \frac{R_\mu}{\lambda_{int}} \simeq A {E_\nu}^n.
$$
Halzen gives $n = 0.8$ and $A = 10^{-6}$ for TeV to PeV 
energies, with $E$ measured in TeV units (\cite{halzen}).

Note here paranthetically that all proposed neutrino detectors will have an
overburden depth less than the radiation length of muons with energies
in the energy range we are considering (see later). 
This means that the above method actually over-estimates the neutrino 
detection probability for downward-going neutrinos 
because the volume of material {\it above} a detector available for the
neutrino to interact within (and subsequently produce a muon which might then
travel on to the detector volume) is substantially less than that for 
upward going neutrinos.

Employing the above (generous) 
parameterization of the neutrino detection probability
allows us to determine 
that the GC signal does not, in fact, emerge clearly from 
the background until well into the upper end of the neutrino energy spectrum
(even for a detector resolution as low as $0.3^\circ$ and the flattest
empirically allowable neutrino spectrum, $\alpha = 2.1$). 
It seems that even granted a detector able to trigger
on neutrino-generated, down-going leptons, 
we cannot, on the basis of our preliminary calculations, confidently conclude
that we might see the GC source in such a way.
We therefore restrict ourselves to consideration of 
the Sgr A East signal to be found in upcoming 
leptons.
    
An immediate consequence of this self-imposed restriction is
that we are unable to reach many conclusions about the usefulness of the
(successor to the) AMANDA neutrino telescope in regards to observing the GC.
This is somewhat unfortunate because the AMANDA experiment, of all neutrino
telescope projects, 
is probably the best currently placed to 
realize the desired $km^3$ status and thus evolve to `IceCube' 
(\cite{icecube}). \footnote{The IceCube project, 
given that it continues to pass through 
scientific review and find funding, will go into construction in the 
2001-2002 Antarctic season and should take 6-7 years to complete. The detector
will be operated as it grows (\cite{halzencomm}).} 
This is because AMANDA/IceCube's
South Polar location means that the GC is always overhead fom the detector.
A detector specific Monte Carlo calculation will probably be needed to settle
whether our particular source can be seen by IceCube.
We can predict, however, that IceCube 
may well be able to
detect the distinctive `double bang' signature of GC $\nu_\tau$ interactions
above the background; see later.  

We must, therefore, look to the Northern Hemisphere for $\nu_\mu$
observing
platforms about which we can make more confident predictions regarding the 
GC source. There are currently four neutrino telescope projects under
development there. The 
Lake Baikal project is a mature experiment, having run on and off since 1993. 
This collaboration has achieved an effective, energy-dependent detector area 
of $1000 - 5000 \ m^2$ and has demonstrated the viability of large-scale,
water-based \v{C}erenkov technology. The collaboration 
is planning for a neutrino telescope of
$5 - 10 \times 10^4 \ m^2$ effective area. This will not 
be a large enough
platform for the relatively high energy (and low flux)
neutrino signal generated at Sgr A East (\cite{baikal}).

Three other projects, all based in the deep Mediterranean, are currently in
the design and prototype stage. They are ANTARES, NESTOR and NEMO. None of 
these projects is guaranteed of the funds to reach $km^3$ status, though
this is the stated goal of all three collaborations. Needless to
say, these deep sea environment projects call for 
great ingenuity and considerable technical innovation. 

The ANTARES
collaboration has completed preliminary reconnaissance of its chosen site at a 
depth of 2400m below the sea near Toulon. They are also well into 
design of electronics and mechanics for the detector. The collaboration's
 current mid-term
goal is to have 13 `strings', with $\sim 1000$ attached PMTs, in place by
2003. Such a configuration would have an effective area of $0.1 \ km^3$
 (\cite{ANTARES}).

The NESTOR collaboration plans for a deployment at $\sim 4000 \ m$ 
depth off the 
south west Grecian coast. This collaboration is at a similar stage of
advancement to the ANTARES group, having completed reconnaissance of their
chosen site and preliminary field testing of crucial components. NESTOR 
also aims
for a $0.1 \ km^2$ effective area detector in the near future
 (\cite{nestor}).

Lastly, the NEMO project is least advanced being in the early R\&D stage. This
collaboration is investigating the suitability of a 
site off the southern Italian coast. They have conducted Monte Carlo
studies of their proposed detector layout (\cite{nemo}).

In conclusion, one does not expect to see a $km^3$ neutrino telescope in the
Northern Hemisphere within a decade, 
but within two decades the chances for such would seem to be quite good.

\subsection{Neutral Current Interactions}

Neutral current (NC) interactions do not 
identify the incoming neutrino flavor and basically constitute a 
background to the more
useful charged current interactions. Energy determination for NC events
is poor because of the missing final state neutrino. Angular determination
is also poor because the single hadronic shower produced is almost point-like
on the scale of a typical detector's PMT spacing. 
NC interactions are only about one 
third as common as CC interactions.  

\subsection{Muon Neutrinos}

The best prospects for observing {\it any} neutrino flux from 
the GC source are offered by muon type neutrinos; $\nu_{\mu}$'s 
and  $\overline \nu_{\mu}$'s (we remind the reader that neutrino telescopes
cannot distinguish a neutrino from an anti-neutrino of the same flavor type).
Charged current interactions of a muon type neutrino
in or near the detector volume result in a nearly
point-like hadronic shower and a high energy muon ($\mu^\pm$)
which, we reiterate, might travel up to ten kilometers and still 
possess enough energy to be detected. Certainly in the above-$100 \ GeV$ 
energy scales of relevance to this paper, muons will be
`uncontained' in the sense that they cannot be expected to be both generated 
and arrested within the $km^3$ detector volumes. Such long tracks mean, of
course, very good determination of the muon's direction of travel. 
On the other hand, the fact that the muons are necessarily uncontained leads 
to uncertainty in energy determination. We now discuss, in the context
of the observation of neutrino-generated muons, the general issues of 
angular and energy determination in more detail.   

\subsubsection{Energy Determination}

An accurate determination of the energy possessed by a muon neutrino (which
produces a muon observed by a detector) is limited by three factors: 
uncertainty in the fraction of the neutrino's total energy 
imparted to the muon, ignorance of the energy loss by the muon outside the
instrumented volume and, finally, the intrinsic 
energy resolution of the detector apparatus
itself (\cite{ANTARES}). 

Regarding the first factor, it can be shown that
the average energy imparted to the muon
is half that of the neutrino in the CC interaction $\nu_{\mu}d \rightarrow 
\mu^-u$ and three quarters in the interaction $\overline{\nu}_{\mu}u 
\rightarrow \mu^+d$  (\cite{ANTARES}). 
A determination of an individual muon's energy, then,
might only give us a minimum energy for the neutrino primary but this 
problem is not a limiting factor if a 
significant number of events 
can be accumulated and we take a statistical view. 

Note that when the muon is uncontained and, hence, an
accurate determination cannot be achieved by measuring the 
length of the entire muon track, a rougher muon energy   
determination can be achieved for $E_{\mu}>1 \ TeV$ 
by measuring ${dE_{\mu}}/{dx}$ because at such energies, where energy
loss is dominated by radiative processes, $({dE_{\mu}}/{dx}) \propto E$. 
It may also eventually 
be possible to glean some neutrino energy information from the 
hadronic shower resulting from the first CC interaction if this happens to
be within the detector volume (keeping in mind the difficulty posed by the
relatively small size of such showers on the scale of 
a next-generation detector's PMT spacing).

That we are dealing with uncontained muon tracks means that one can only 
arrive at a minimum original muon energy. That we can
make some sort of energy determination from ${dE_{\mu}}/{dx}$, though,
means that we have a much better idea of the original energy of a 
{\it totally} uncontained muon 
%(i.e. one that is neither stopped not `started'
%within the detector volume) 
than would be imparted by just assigning it a 
minimum energy enough to take it across the detector.

Given all the above factors, the ANTARES collaboration has judged
on the basis of Monte Carlo simulations of their detector array that they 
can gauge a muon neutrino's energy to within a factor of three for 
$E_{\nu} > 1 \ TeV$ (\cite{ANTARES}).

\subsubsection{Angular Determination}

Again three factors limit the determination of the primary neutrino's
direction of travel. These are the uncertainty in the angle between the
incoming $\nu_{\mu}$ and the resulting $\mu$, the deviation 
of the $\mu$ away from its original direction of travel due to 
multiple scattering and, lastly, the detector's intrinsic angular resolution
as determined by uncertainties in its exact geometry, etc. (\cite{ANTARES}).
Of course, the severity of the first two 
problems decreases with increasing energy, but the relative severity of the
two likely changes with energy. For example, the ANTARES collaboration has
determined from MC simulations that below $10 \ TeV$ total angular resolution
is limited by detector effects whereas above $100 \ TeV$ it is limited by 
the unavoidable angular distribution of the neutrino interactions. They claim
an angular resolution of $0.3^{\circ}$ is achievable (\cite{ANTARES}).
With such a 
resolution the GC signal is above atmospheric neutrino background for 
energies greater than a few $\times \ 100 \ GeV$. 

The AMANDA project (which will hopefully evolve into IceCube) 
has to contend with the short scattering length of the \v{C}erenkov
light in ice, $>200m$, as compared to seawater at $24m$. 
Despite this, IceCube will achieve an angular resolution less than one 
degree and perhaps as low as 
$0.4^{\circ}$ (\cite{halzencomm}). We note parenthetically that 
such a resolution will mean that many
southern sky sources (i.e. sources of downgoing neutrinos)
{\it will} be able to be 
seen by IceCube above atmospheric muon background.

\subsubsection{Earth Opacity}

At $\sim 4 \times 10^{13} \ eV = 40 \ TeV$ 
the interaction lengths of all neutrino flavors
become less than the Earth's diameter. This means that, in particular, 
$\nu_\mu$'s
are unlikely to reach a detector from a nadir angle of $0^\circ$
 (\cite{ANTARES}). (The same is true for $\nu_e$'s but {\it not} $\nu_\tau$'s;
see later.) The 
attenuation of the $\nu_\mu$ interaction length continues until at
$\sim 10^{15} \ eV$ it is less than a very small fraction of the Earth's 
diameter, so that this flavor is attenuated over all nadir
angles, even those approaching the horizontal. At such high energies, then, 
the Earth is said to be `opaque' to  $\nu_\mu$'s (and $\nu_e$'s) (\cite{nico}).
We must take both this effect and our self-imposed requirement that 
the GC neutrino
source be below the horizon from the observation point
(in order to avoid the atmospheric muon background problem)
into account to generate a more realistic
estimate of the event rate due to the GC source. 

Let us assume the best case scenario for $\nu_\mu$ fluxes -- 
detector angular resolution of $0.3^{\circ}$, 
and a neutrino spectrum that goes as $\alpha  =  2.1$ -- to make 
a determination of the expected event rate in a hypothetical,
$km^3$ detector located on the proposed ANTARES site.
Note that with this revised angular resolution, the GC neutrino flux
is above atmospheric neutrino background at an energy around 
an order of magnitude 
lower than previously: a few $100 \ GeV$.
Also note that
the GC is below the horizon about two thirds of the time from 
this latitude (\cite{zombeck}) and, therefore, 
invisible at least one third of the time (even if low enough detector
resolution were
 achieved to unequivically avoid the atmospheric muon background 
problem, ANTARES is being designed with downward pointing PMTs). 
Adopting the neutrino
penetration coefficients calculated by Naumov and 
Perrone (\cite[fig.3]{naumov}), we
determine that the expected annual 
event rate from $\nu_\mu$'s generated at the GC
is $\sim 40$ for the
no-oscillation case and $\sim 20$ if oscillations do occur. For  
$\alpha  =  2.4$, but retaining an angular resolution of $0.3^{\circ}$, 
we expect
$\sim 5$ events without oscillation and $\sim 2$ with. Clearly, then,
we approach the lower end of statistical relevance with this value for 
$\alpha$. (In these calculations we have not allowed for the regeneration
effect due to NC interactions that affects all neutrino flavors. We expect
this effect to be small (\cite{martin}).)

\subsubsection{Muon Neutrino Background}

We note, in passing, one unavoidable source of $\nu_\mu$ background;
CC $\nu_\tau$ interactions can mimic CC $\nu_\mu$ interactions if 1) the 
$\nu_\tau$ energy is too low to effectively separate the original
CC interaction vertex and the $\tau$ decay vertex, and 2) the  $\tau$ decays
muonically (the branching ratio for this decay is $\sim 17$ \%  (\cite{caso})).

\subsection{Electron Neutrinos}

In contrast to the case for muon neutrinos, the prospects for 
identifying electrons ($e^{\pm}$)
 in a detector generated by $\nu_{e}$'s from the 
direction of the GC seem remote. Quite a few of the significant problems
with observing the $\nu_e$ signal can be related back to
the relatively tiny propagation length ($\sim$ meter) of 
high energy electrons (and positrons) 
in matter. Perhaps most significant is that, 
as with NC interactions,
the hadronic and electromagnetic showers initiated by a  $\nu_{e}$ in a CC
interaction have almost point-like dimension on the scale of the proposed
detectors and, hence, provide little directional information. Thus, even if
we grant that an electron signal in the appropriate 
energy range for the GC source might be identified, we cannot actually 
identify the origin of the primary electron neutrino. 

A second problem is that electron 
neutrino initiated CC events are very difficult to
conclusively identify. In principle a smoking gun for such events 
is presented by the coincident 
presence of both a hadronic shower (from the disturbed
nucleus) {\it and} an electromagnetic shower from the quickly braked $e^\pm$.
It is very difficult, however, for the proposed, next-generation 
 \v{C}erenkov technology to distinguish between the
two types of showers. Both showers, 
we repeat, are essentially point-like on the
scale of the typical detector's PMT spacing and, after all, are 
observed only indirectly through the \v{C}erenkov flash they produce. Thus, NC
events, which produce a point-like hadronic shower, are difficult 
to distinguish from CC $\nu_e$ initiated events and provide a significant
background problem. Further, even imagining that we had some reliable
technology to identify the presence of a high energy electron,
the CC interactions of $\nu_\tau$'s can still mimic 
CC $\nu_e$ events if 1) the $\tau$ energy is not high enough to ensure that
the hadronic shower from the CC interaction of the primary $\nu_{\tau}$ 
and the later decay are effectively separated on the scale of the
detector, and 2) the $\tau$ decays electronically (with a branching ratio
of $\sim18$ \% (\cite{caso})). 

A yet further problem is the fact that the short path of the electron in matter
means that one can only register contained $\nu_e$ CC events, dramatically
reducing the effective volume monitored by the detector in comparison with
$\nu_\mu$ events.

Altogether one cannot but conclude that the chances for detecting GC 
$\nu_e$'s, at this stage, seem remote.

\subsection{Tauon Neutrinos}

Although the chances for observing GC $\nu_\tau$'s seem more hopeful than 
those for GC $\nu_e$'s, there will still be considerable problems with
this flavor. At least two unique signatures for the $\nu_\tau$ have
been identified in the literature: 1) the `double bang' and 2) flat angular
dependence of the signal or `pile up' (\cite{nico}; \cite{learned}; 
\cite{saltzberg}; \cite{iyer}).
 These both, however,
 tend to become significant on the 
higher energy side of the GC neutrino spectrum. 

\subsubsection{Double Bang}

In more detail, the `double
bang' signal requires that a $\nu_\tau$ 
undergo a CC interaction in the
detector volume to produce a $\tau$. If the energy of this $\tau$ is high
enough then the 
hadronic shower resulting from the initial interaction
of the neutrino primary and the later hadronic shower resulting from the 
$\tau$ decay will be resolvable on the scale of the detector. Exactly
where the resolvability threshold is can probably only be determined by
detector-specific MC simulations. The ANTARES group believes the signal 
certainly {\it cannot} be resolved 
for $E_{\nu_\tau} < 100 \ TeV$ (\cite{ANTARES}).
At $E_\tau \sim PeV$, towards the upper limit of the GC neutrino spectrum,
 the two bangs should be separated by about $100m$ and clearly resolvable. 

The usefulness of this
signature, then, will depend on detector specifics and the question of whether
a statistically significant flux can be obtained from whatever part 
of the $\nu_\tau$ spectrum remains able to produce a signal. 

One also notes that Earth opacity will significantly
reduce the flux of $\nu_\tau$'s 
sufficiently energetic to produce the double bang signal if one is looking
for the signal in upcoming neutrinos. It is in searching for the double bang
signature from GC $\nu_\tau$'s, then, 
that we can predict that 
AMANDA (or, more precisely, IceCube the $km^3$ extension of AMANDA)
may well find employment in
regards to this source; GC neutrinos will not be affected by Earth opacity
when observed by IceCube. (The genuine GC $\nu_\tau$ flux is
substantially above that of the atmospheric $\nu_\tau$'s due to `prompt'
and conventional flux over an angular resolution even as bad as $2^\circ$
 (\cite{pasquali}) and $2^\circ$ is a pessimistic prediction for 
the IceCube's angular resolution (\cite{halzencomm,halzen})).
Assuming a best-case scenario for $\nu_\tau$ detection, viz,
 the flattest allowable
GC spectrum ($\alpha = 2.1$), double bang resolvability all the way
down to $100 \ TeV$ and $\nu_\mu \rightarrow \nu_\tau$ oscillations, and
assuming a double bang detection probability given by 
${1 \ kmwe}/{\lambda_{int}}$ ($1 \ kmwe$ means $1 \ km$ water equivalent),
we can arrive at an (optimistic) annual event rate prediction for 
IceCube. 

We derive the double bang detection probability by employing similar
logic to that which led to the $\nu_\mu$
detection probability presented previously. 
The difference here is that we assume the $\tau$ decay length is small on the
scale of the linear dimension of the 
detector (hence the 1 in the numerator), 
whereas previously we assumed that the $\mu$ radiation length is
large in comparison to this scale. We employ a parameterization 
of the neutrino interaction length presented in graphic form (fig. 11)
 in  (\cite{gandhi1}).

Using the detection probability described above, and the best case scenarios
for the GC spectrum and double bang resolvability, we
determine an event rate of 1 double bang signal per year.
This is at the threshold of detectability.

%We note that
%that the $\nu_{\mu}$ event rate for  $E_{\nu_\mu} > 500 \ TeV$ with
%oscillations and
%for a best case scenario -- a spectrum that goes as $\alpha = 2.1$ 
% -- in a $km^3$
%detector might reach $\sim 20$ events per year. We would expect the
%double bang event rate to be substantially smaller than this given that
%only fully contained events can generate a signal. 

\subsubsection{Pile Up}

The idea behind the second $\nu_\tau$ signature 
-- the flat angular dependence which has recently received attention 
from  Halzen and Saltzberg (1998) and Iyer, Reno, and Sarcevic (1999)
-- is to actually make positive use of the Earth opacity previously mentioned.
When $E_{\nu_\tau}$ climbs beyond $\sim 4\times10^{13} \ eV$
the interaction length
of the  $\nu_\tau$ becomes, as for the $\nu_e$ and $\nu_\mu$,
less than the Earth radius. But
whereas $e$'s and $\mu$'s resulting from CC interactions are stopped in the
Earth, $\tau$'s
from CC interactions decay back to $\nu_\tau$'s before being stopped,
 producing a neutrino with
something around one quarter the energy of the original and traveling 
in much the same direction. This process can occur more than once, 
each iteration producing a progressively lower energy $\nu_\tau$, ensuring 
that whatever the energy of the primary $\nu_\tau$, a $\nu_\tau$ signal from
a point source
should reach a detector on the other side of the Earth. This signal will
exhibit a `pile up' just below the energy where the $\nu_\tau$'s interaction 
length becomes greater than the fraction of the Earth's diameter subtended
by a ray from the source to the detector.

In other words, for $\nu_{e,\mu}$ energies in excess of $\sim 10^{12} \ eV$, 
as the angle between a neutrino source and the nadir is decreased from 
$90^\circ$, a critical angle will be reached where the 
$\nu_{e,\mu}$ flux will begin to be attenuated. This attenuation increases
to reach a maximum at $0^\circ$. Further, as the energy of the $\nu_{e,\mu}$
signal increases, the flux attenuation sets in at increasingly large
(i.e., increasingly horizontal) angles.

On the
other hand, the $\nu_\tau$ flux, although shifted downward in energy, 
should still be the same. This results in
 the flat angular dependence of the $\nu_\tau$ 
part of the signal at high energies and, given a significant 
$\nu_\tau$ component of the total
neutrino flux, a flatter than expected angular dependence of the total neutrino
flux. 

One way to search for a $\nu_\tau$ signal, then, is through the decay chain
$\tau \rightarrow \nu_\tau \mu \nu_\mu$ (branching ratio
$\sim 17 \%$ (\cite{caso})). 
Given the above considerations, if we assume that a
significant part of a neutrino signal is due to $\nu_\tau$'s, we expect an
enhancement of the number of $\mu$'s coming from the direction of our source,
below certain energies and nadir angles, over that expected from the `raw'
$\nu_\mu$ and $\nu_\tau$ fluxes. In order to see this enhancement, however,
we require that the $\nu$ energy spectrum not be too steep. 
Otherwise the increase
of the $\mu$ flux in some particular, lower energy `bin' will be insignificant
on the scale of the number of events that would be recorded there anyway due
to the raw $\nu_\mu$ and $\nu_\tau$ fluxes.

Iyer, Reno, and Sarcevic (199) have made calculations 
of the `pile up' enhancement
for neutrino spectra which go as different negative powers: $n=1,2,3.6$. 
For $n=1$ the enhancement is a noticeable effect, but for
$n=2$ and greater the spectra are too steep for the effect to be
discernible. For the Sgr A East neutrino flux, with a best-case spectrum which
has an $n$ of 2.1, we must unfortunately conclude that the above diagnostic
for the presence of a significant $\nu_\tau$ component in the total neutrino
signal will not be useful.

In summary for the $\nu_\tau$ case, we believe that the GC can produce
$\nu_\tau$'s energetic enough to produce a double bang signal, but that
the spectrum is too steep to evidence $\nu_\tau$'s with pile up. 
A preliminary calculation reveals a double bang signal at the threshold 
of detectability in IceCube, but a confident indication that this
signal will produce a statistically significant event rate 
requires a detector-specific study.

\subsubsection{Tauon Neutrino Background}

As has been mentioned, we expect no $\nu_\tau$ flux from
pion decay from p-p scattering at the 
GC in the absence of oscillations and, hence, observation of 
a $\nu_\tau$ flux of the order of the $\nu_e$ or $\nu_\mu$ 
flux constitutes {\it prima facie} evidence
for exactly such neutrino oscillations. One must be concerned, however, about
sources of background to the  $\nu_\tau$ oscillation signal, both genuine 
$\nu_\tau$ flux from sources that have not been accounted for and false 
$\nu_\tau$ signals in the detector. 

One source of $\nu_\tau$'s that we can anticipate at higher energies at 
the production site is the decay of charmed mesons (principally $D_s$) produced
in p-p scattering through $\tau$ and  $\nu_\tau$ production. It should be
noted that the cross-sections for $c$ and ${\overline c}$ production via
p-p scattering are greatly uncertain in the energy range of interest, as are
the fractional likelihood of $c \rightarrow D_s$ and the branching ratio
for $D_s \rightarrow \tau \nu_\tau$ (\cite{caso}; \cite{pasquali}). 
In comparison, however, with pion production 
processes leading to $\nu_{e,\mu}$ such charmed meson production is still
greatly suppressed. The flux ratio 
${{\Phi_{\nu_\tau}}^{obs}}/{{\Phi_{\nu_\mu}}^{obs}}$
can still, therefore, be expected to be a small number taking this process into
account, although there might be considerable deviation from $1$ in 
${{\Phi_{\nu_\tau}}^{obs}}/{{\Phi_{\nu_\tau}}^{theor}}$ 
(if we assume large statistics) 
without oscillations necessarily being implied. 

\section{Observational Consequences -- in Practice}

If we grant that the GC source will not produce $\nu_e$'s in 
an observational energy range, might produce $\nu_{\tau}$'s in an observational
range and certainly will produce observational $\nu_{\mu}$'s, we are left with
only one useful flux ratio that is certainly measurable:
\begin{equation}
\frac{\Phi_{\nu_{\mu}}^{obs}}{\Phi_{\nu_{\mu}}^{theor}},
\end{equation}
and two that may
be measurable:
\begin{equation}
\frac{\Phi_{\nu_{\tau}}^{obs}}{\Phi_{\nu_{\mu}}^{obs}}\\, \qquad  \qquad 
\frac{{(\Phi_{\nu_{\tau}}}/{\Phi_{\nu_{\mu}})^{obs}}}
{({\Phi_{\nu_{\tau}}}/{\Phi_{\nu_{\mu}}})^{theor}}\\.
\end{equation}

As 
previously discussed, deviation from one in the first ratio, 
by itself, would provide only weak evidence for oscillations
   unless the empirical values of $\alpha$ and $\Phi_{\nu}(10 \ GeV)$
 were further constrained (by future $\gamma$-ray observations). 
Even if this were achieved, however,
given the indirectness of the 
$\Phi_{\nu_{\mu}}^{theor}$ measurement, there would have to be some
doubt about whether the presence of oscillations had been
conclusively demonstrated.  
With empirical determination of the values of 
all three ratios, only scenario 3 emerges with a 
unique signature. Otherwise, we can only distinguish the $\nu_\mu \to \nu_s$
scenarios (1 and 4)
from the  $\nu_\mu \to \nu_\tau$ scenarios (2 and 5), 
without being able to
conclude anything about $\nu_e$ mixing. Certainly, however, 
${\Phi_{\nu_{\tau}}^{obs}}/{\Phi_{\nu_{\mu}}^{obs}}$ potentially offers
very strong evidence of oscillations if it is found to deviate substantially
from zero.

Given that the measurement 
of these ratios lies at least a decade
into the future it is, in fact, not unlikely that the uncertainty  
regarding the $\nu_e$ and 
$\nu_\mu$ oscillation modes be largely dispelled by the time of 
such measurement, i.e., other
experiments will determine which of scenarios 1 to 5 (or
bimaximal or trimaximal oscillations or even one of the non-oscillation
scenarios -- see below) actually occurs in nature.
The most interesting science that might be
extracted from GC neutrino observations, then, 
may be an empirical determination of $\alpha$ and $\Phi_{\nu}(10 \ GeV)$
independent of $\gamma$-ray observations.\footnote{With the 
certain knowledge that $\nu_\mu \to \nu_x$ oscillations {\it do}
take place and, hence, the knowledge that
${\Phi_{\nu_{\mu}}^{obs}}/{\Phi_{\nu_{\mu}}^{theor}}$ must be
${1}/{2}$, and from an empirical determination of the
$\nu_{\mu}$ spectrum, one can work backwards to obtain
$\Phi_{\nu_{\mu}}^{theor}$ and, thence, $\alpha$ and $\Phi_{\nu}(10 \ GeV)$.} 

Note in passing that determination of the flavor composition
of the GC neutrino signal could certainly
provide for stringent tests of various 
alternative, no-oscillation explanations to the solar and atmospheric 
neutrino anomalies if these are not ruled out in the near future. 
For instance, a large $\nu_\mu$ component in the GC neutrino
spectrum would imply a 
much larger lower limit on the `$\nu_\mu$ lifetime' (we should, strictly,
consider the lifetimes of the mass eigenstates composing the $\nu_\mu$)
than is required to explain the atmospheric neutrino anomaly. See, e.g.,
 (\cite{barglearn}). 
On the other hand, 
%a $\nu_\mu$ component substantially 
%smaller than predicted by a no-oscillation calculation would rule out
flavor changing neutral currents (FCNC), invoked as explanations of the
atmospheric anomaly (\cite{fcnc1}; \cite{fcnc3}; 
\cite{fcnc4}; \cite{fcnc5}), cannot affect 
the Sgr A East signal. This is because the column density encountered
by neutrinos propagating from the GC to the Earth is far too small 
to allow this mechanism to occur. 
FCNC explanations of the atmospheric anomaly,
then, predict a GC neutrino event rate undiminished from the na\"{\i}ve
expectation and deviation from this would tell against such explanations. 

\section{General Background Problems}

There are a number of sources of background to the GC neutrino signal. 
Logically, we can break these down into the two general classes: 1) `enshrouded
sources' and 2) terrestrial background. By the former we refer to any sources 
of genuine neutrino signal from the GC which are `hidden' in the sense that
they are not correlated with the GC $\gamma$-ray spectrum. 
By the latter we mean
the atmospheric neutrino and muon backgrounds that are endemic. These two
have already been addressed.

\subsection{Background from Enshrouded Sources}

We know of two potential sources of an enshrouded neutrino signal from the
GC. One -- neutrino production via high energy cosmic ray scattering on
the ambient material in the Galactic plane -- is virtually assured
 (\cite{gaisser}; \cite{ingelman1}). The other
-- neutrino production via annihilation of WIMPs accumulated 
in the gravitational well at the GC -- is a possibility
 (\cite{jungman}; \cite{silk}).

\subsubsection{Neutrino Production off the Interstellar Medium}

Note that the first background 
source is, like the Sgr A East source, due to decay
of pions produced in nucleon-proton scattering. The density of ambient matter
in the galaxy is greatest, in general, in the Galactic plane and greatest of
all at the GC so we may expect a large background neutrino flux from this
direction. Of course, the pionic decay also leads to the production of 
$\gamma$'s. That we consider this neutrino source enshrouded, then, 
is due to the relatively large angular resolution of the proposed neutrino
telescopes; the neutrino telescopes see neutrinos from a much larger area of
sky than the $\gamma$-defined size of Sgr A East. 

Detailed estimates have been made of the rate of neutrino production 
by the interaction of cosmic rays with the interstellar medium 
 (\cite{ingelman1}). 
This neutrino flux has been shown, however, to be below the
atmospheric neutrino background for much of the energy range
 under consideration. Even given that the GC background 
exceeds the atmospheric one above $\sim 5 \times 10^{14} \ eV$, the background
from $0.3^{\circ}$ of sky (as relevant for ANTARES) 
is still considerably below the signal.

\subsubsection{Neutrino Production from WIMP annihilation}    

The exact flavor composition of the neutrino flux generated by
WIMP annihilation is model-dependent. It is
conceivable, for instance, that a large $\nu_\tau$ component might be present
in this signal, if it exists at all. There is a fairly robust and 
model-independent upper bound to
the WIMP mass of $300 \ TeV$ (\cite{jungman}). 
Neutrinos
generated in WIMP annihilation processes will have typically between
one half to one third the WIMP rest mass energy  (\cite{ANTARES}). We cannot, 
therefore, strictly rule out the possibility that the Sgr A East neutrino 
signal is polluted with neutrinos from WIMP annihilation. We do not consider
this possibility in any detail, however, because, most reasonable WIMP 
candidates have maximum masses some orders of magnitude below this. The 
neutralino, for instance, cannot be more massive than $\sim 3 \ TeV$ if it
is to be a WIMP candidate  (\cite{jungman}). Neutrinos produced in its
decay, therefore, can, at worst, be just below the energy cut-off of the part
of the GC neutrino signal we are examining.
 
\section{Conclusion}

The GC neutrino source should produce an observable 
oscillation signature. The strongest evidence
for such would take the form
of a $\nu_\tau$ flux attaining a significant fraction of 
the $\nu_\mu$ flux from from this source.
Such a $\nu_\tau$ flux may be inferred from the 
double bang signature at IceCube, the $km^3$
successor to the AMANDA
telescope. Detector-specific simulations are required 
for a confident determination of whether
the double bang event rate due to the GC will be statistically significant in
the event that either of scenarios 2, 3 or 5 is correct, but
preliminary calculations reveal that this event rate may be just at the
threshold of detectability. Such simulations are also required to determine
whether IceCube might see the GC $\nu_\mu$ signal
 against the atmospheric {\it muon}
background.

A deviation from the expected $\nu_\mu$ flux
determined from $\gamma$-ray observations of the GC is guaranteed for all
neutrino oscillation scenarios identified. Observation of such 
deviation would, however,
constitute more equivocal evidence for oscillation than a strong
$\nu_\tau$ signal because of uncertainties in the total expected
neutrino flux calculated on the basis of $\gamma$-ray observations.
Certainly, the value of $\alpha$, the 
numerical power of the power-law proton spectrum at Sgr A East,
would have to be further constrained before the above became a useful
diagnostic (as would $\Phi_{\nu}(10 \ GeV)$). 
The actual $\nu_\mu$ flux should be able to 
be inferred from the $\nu_\mu$ event rate experienced by a future,
Mediterranean-based $km^3$ \v{C}erenkov neutrino detector.

%Observation of an oscillation signature will not, unfortunately,
% further constrain
%the oscillation parameter space already mapped out by atmospheric, solar,
%reactor and accelerator oscillation experiments.

Strong confirmation of the oscillation signature will require
observation of  $\nu_e$ flux from the GC to see whether 
the $\nu_e$ to $\nu_\mu$ ratio varies significantly from ${1}/{2}$
(though, as discussed, if the
small mixing angle solution to the solar neutrino problem is correct
$\nu_e$'s will not oscillate on their way from the GC). 
The energetics of the GC `beam', 
however, place it below the region where next-generation techniques and
detectors are currently predicted to be able to identify a $\nu_e$ component.
Such confirmation, then, must lie some decades into the future.

Perhaps the best science that might be extracted from the GC neutrino
spectrum as observed by a future $km^3$ \v{C}erenkov
neutrino detector, assuming that other experiments resolve
the electron- and muon-type neutrino oscillation mode questions first,
is an empirical determination of $\alpha$.
By such a determination, a neutrino telescope would realize
the aspiration expressed in its very name, that, at base, it 
is a device for investigating the nature of astronomical objects, not merely 
the radiation they emit.\footnote{A similar point is made by 
Raffelt (1998).}

\section{Acknowledgments}

R.M.C. gratefully acknowledges useful discussion and 
correspondence with D. Fargion, F. Halzen, A. Oshlack, W. Rhodes, 
and M. Whiting. R.R.V. would like to thank V. Barger for a useful
discussion; he is supported by the Australian Research Council. R.M.C. is
supported by the Commonwealth of Australia. F.M. is partially supported by NASA
under grant NAGW-2518 at The University of Arizona.

%\begin{references}

%\end{references}

\end{document}